\documentclass[10pt,twocolumn]{IEEEtran}
\usepackage{multicol}

\usepackage{color}
\usepackage{xcolor}
\usepackage{soul}
\usepackage{amsmath}
\usepackage{graphics}
\usepackage{setspace}
\usepackage{cite}
\usepackage{latexsym}
\usepackage{float}
\usepackage{epsfig}
\usepackage{multirow}
\usepackage{cite,cases,url}
\usepackage{amssymb}
\usepackage{graphicx}
\usepackage{epstopdf}
\usepackage{balance}
\usepackage{subcaption}
\usepackage{enumerate}
\usepackage{array}
\usepackage[ruled,linesnumbered]{algorithm2e}
\usepackage{enumitem}
\usepackage{caption}
\usepackage{subcaption}
\usepackage{tikz}
\usepackage{adjustbox}
\usepackage{etoolbox}
\usepackage[noend]{algpseudocode}
\makeatletter
\usepackage{tikzpagenodes}      
\usepackage[normalem]{ulem}
\useunder{\uline}{\ul}{}

\newcolumntype{L}{>{\centering\arraybackslash}m{5cm}}
\newcolumntype{K}{>{\centering\arraybackslash}m{6cm}}
\newcolumntype{P}{>{\centering\arraybackslash}m{2.3cm}}
\newcolumntype{M}{>{\raggedright\arraybackslash}m{2cm}}
\newcolumntype{N}{>{\raggedright\arraybackslash}m{2.5cm}}

\usepackage{changepage}
\algrenewcommand\algorithmicindent{12.0em}%
\usepackage{xpatch}
\makeatletter
\newcommand{\removelatexerror}{\let\@latex@error\@gobble}
\makeatother

\begin{document}

\title{Securing Mobile Multiuser Transmissions with UAVs in the Presence of Multiple %\hl{Smart} 
Eavesdroppers}

\author{\IEEEauthorblockN {Aly Sabri Abdalla %\IEEEauthorrefmark{1}
and
Vuk Marojevic%\IEEEauthorrefmark{1}
}

\IEEEauthorblockA{%\IEEEauthorrefmark{1}
Department of Electrical and Computer Engineering, Mississippi State University, MS 39762, USA\\}

Email: $\{$asa298$|$vm602$\}$@msstate.edu%, vuk.marojevic@msstate.edu
\vspace{-4 mm}
}

\vspace{-4 mm}

\maketitle
\begin{tikzpicture}[remember picture,overlay]
   
    \node[align=center,text=black] at ([yshift=-2em]current page text area.south) {
    \footnotesize Manuscript is published in: IEEE Transactions on Vehicular Technology ( Volume: 70, Issue: 10, Oct. 2021). Date of publication 10 August 2021. \\[-0.3 em] \footnotesize This work is supported in part by the NSF PAWR program, under grant number CNS-1939334.};
  \end{tikzpicture}%

\begin{abstract}
This paper analyzes the problem of securing the %uplink 
transmissions of multiple mobile users against %multiple \hl{smart}
eavesdropping attacks. We propose and optimize the deployment of a single unmanned aerial vehicle (UAV), which serves as an aerial relay (AR) between the user cluster and the base station. The focus is on maximizing the secrecy energy efficiency %(SEE)
by jointly optimizing the uplink transmission powers of the ground users and the position of the UAV. The joint optimization problem is nonconvex; therefore we split it into two subproblems and solve them using an iterative algorithm.  

\textit{Index Words}--AR, cellular communications, secrecy rate, security, UAV.

%. Modeling and estimation ofmmWave blockage remain at the heart of the challenges behindits deployment especially in urban areas. In this paper, we
\end{abstract}

\IEEEpeerreviewmaketitle

\vspace{-5 mm}
\section{Introduction}
\label{sec:intro}

%% Vuk: using %% on Aug. 7

%%5G and beyond wireless networks will offer a variety of services and enable new  %varieties of new services and 
%%applications, including mission-critical data and control signaling. % that required a specific quality of service (QoS), reliability, low latency, and enabling enormous connectivity of devices . Some examples of those services are massive machine type communication (mMTC), ultra-reliable low latency communication (URLLC), and vehicle-to-everything (V2X)~\cite{5G}. 

The worldwide deployment of 4G long-term evolution (LTE) and now 5G New Radio (NR) networks and the increasing number of vertical industries that those networks support require secure and reliable %, and privacy preserving
communications. 
%However, these ostentatious technologies and services encounter a wide range of security and privacy difficulties, which have not been fully investigated~\cite{5GSecurity}. 
Unmanned aerial vehicles (UAVs) %have been emerging in the commercial spaces technology that 
will be fully integrated into cellular networks and will be able to assist 5G and beyond networks~\cite{Aly2021std}. %~\cite{UAV5G_Recent}~\cite{UAV5G_IoT}. 
UAVs come in many forms. They have flexible mobility patterns, can hover, be tethered, and establish line of sight (LoS) connectivity with ground users at low deployment costs. An emerging application for UAVs %that is currently in high demand 
is the use of rapidly deployable aerial support nodes that can enhance the %physical layer
security of terrestrial or aerial wireless networks~\cite{aly202WCM}.
 
This paper focuses on the security of the cellular radio access network %(RAN) 
and, in particular, the privacy of broadband %user data. 
transmissions. 
We analyze the feasibility of using a single UAV to improve the security of several ground cellular network users against eavesdropping. 
%5G wireless networks physical layer security 
%Specifically we focus on the eavesdropping attack.
Eavesdropping is a fundamental attack, where the eavesdropper is a radio receiver that illegitimately captures information transmitted over the air. %the transferred  between nodes. %passively.
This attack can %directly
compromise the privacy of sensitive data, missions, or systems. 
%When able to capture signals and messages, 
It can also facilitate effective follow-up attacks %can be successively launched
on the data integrity or service availability, among others. %of and  the system where the attacker can wiretap the transferred sensitive legitimate information between nodes passively. 
 
%A number of existing research items have studied the use of 
UAVs have been considered as wireless support nodes for improving the physical layer security. %~\cite{UAVAly}. 
This includes the deployment of aerial relays (ARs) and jammers for protecting a single terrestrial link against a ground or aerial eavesdropper~\cite{AlyVTC20}. Reference~\cite{PoWTraj1} aims at protecting a single UAV-to-ground transmission against %multiple fixed
static eavesdropping attacks by optimizing the UAV trajectory and transmission power of the UAV. The authors of~\cite{PoWTraj2} propose a similar technique to enhance the secrecy rate of UAV-to-ground communications under a single static eavesdropping attack at a known location. The work presented in~\cite{Jamm1} secures the downlink (DL) transmission from an aerial base station (BS) %mounted on an UAV 
to multiple %fixed 
ground users in a time division multiple access manner while another UAV is jamming the signals. It %employs %through 
%a 
jointly optimizes the user scheduling, transmission powers, and the UAV trajectory. Similarly,~\cite{Jamm2} designs an %relative
optimization problem for the same environment, but for frequency division multiple access.       
%This includes solutions that jointly optimize the UAV trajectory and transmit power~\cite{PoWTraj1,PoWTraj2} and solutions that jointly optimize the user scheduling, trajectory and transmit power for cooperative friendly jamming from the UAVs~\cite{Jamm1,Jamm2}. 
%\textcolor{red}{\hl{Discuss more about these works: context, solution and main result/conclusion.}}
 
Prior research has considered the UAV as an AR %problem of 
for securing a single link %of a single static user
against eavesdropping. Mobile and vehicular users are assumed in~\cite{AlyIoT} and \cite{bodong19}, which explore the benefits of a single AR per user. Such solutions are not scalable. References~\cite{Jamm1} and \cite{Jamm2} address the multiuser scenario, but with static users. %To the best of our knowledge, no research have been
%To the best of our %authors' knowledge, t
There is a research gap in analyzing the %none of the relevant literature has studied the 
scalability performance of an AR supporting multiple mobile %ground
users against multiple eavesdroppers. 
%%We %develop  to
%%thus investigate the performance of an AR as a mitigation technique against %smart
%%eavesdropping attacks targeting a cluster of mobile ground users. 
%More precisely, 
This paper therefore proposes and analyzes a practical %joint 
UAV positioning and multiuser power control (PC) scheme to maximize the secrecy energy efficiency (SEE) for efficient and secure multiuser %uplink 
broadband communications. Our results show that a single AR can effectively improve the SEE of a cluster of mobile users. %transmissions. from the cluster of mobile ground users. 
%{\textcolor{red}{\hl{Is it a mobile cluster or a cluster of mobile users?}} \textcolor{blue}{I think it is more a mobiel cluster because all users move with same speed and for the same direction}
% The authors of~\cite{UAVSafeguarding} discusses the new security challenges that can effect the aerial and ground wireless networks with the deployment of the UAVs. After that, they studied some of solutions to tackle these challenges including the use of UAVs as an aerial relay or aerial jammer specifically for the ground eavesdropping attack.Also, some of the prior works presented in~\cite{PoWTraj1,PoWTraj2} studied the joint problem of transmit power and trajectory to improve the physical layer security of the system. 
 
We present the system model, problem formulation and our iterative solution in Sections II, III, and IV. Section V provides the numerical results. Section VI derives the conclusions.

\vspace{-2mm}
\section{System Model} 
\vspace{-1 mm}
\label{sec:system}
% ...

% ...

% <SYSTEM MODEL THAT IS THE BASIS OF OUR NUMERICAL ANALYSIS (SIMULATOR)--START SIMPLE, WE WILL ELABORATE LATER>

% ...
 \begin{figure}[t]
     \centering
     \includegraphics[width=0.385\textwidth]{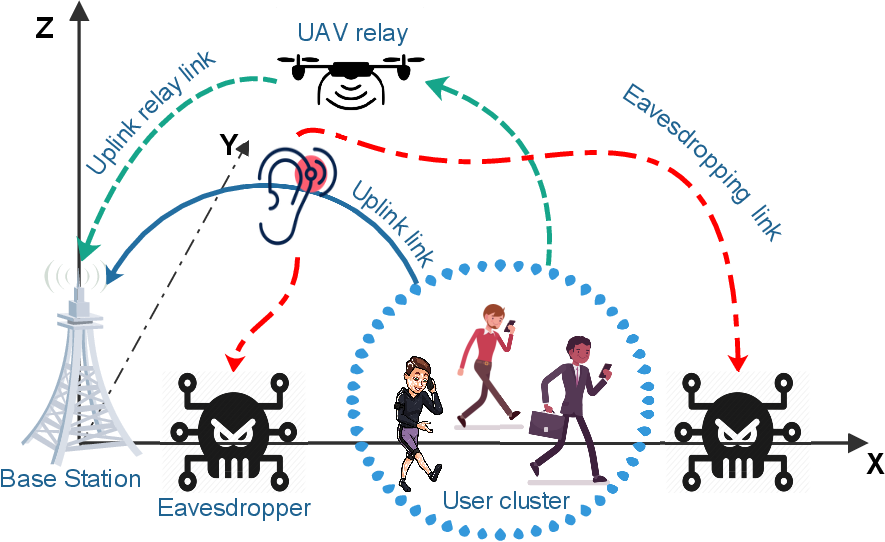}
     \caption{Multiuser eavesdropping scenario.}
     \vspace{-5mm}
     \label{fig:system}
 \end{figure}
 
% \subsection{System Model}
%\subsubsection{Sub-problem (P2.I): Transmission Power Optimization Under Provided Trajectory}
Fig.~\ref{fig:system} illustrates the scenario considered in this work. A cluster of mobile ground users communicates with a BS %transmits information to the base station as an uplink transmission with 
in the presence of terrestrial %smart
eavesdroppers. 
%%We assume that the user cluster moves along a path and %s along the x axis %with fixed a step 
%%encounters %while the 
%%eavesdroppers, which are static here. 
The active users in the cluster %has a predefined number of users and their locations 
are randomly distributed. %within the radius of the cluster.
%\hl{The eavesdropper nodes will apply a smart selection wiretapping attack %to preserve their resources as possible 
%by attacking the closet and most vulnerable users with respect to their location. 
%This strategy can ensure a strong signal-to-noise ratio (SNR) of the link between the victim node and the eavesdropper.} %and the selected .
The UAV is positioned to support the users %under attack 
by relaying their %uplink
data streams. % to the BS. %\hl{See my comment}
Without loss of generality, in this paper we model and analyze the uplink (UL) channel. The same principles can be applied for the DL.

%For notational convenience, 
The set of $M$ users in the cluster %being attacked 
and the set of $K$ eavesdroppers are denoted as $V=\{V_i|i=1,2,...,M\}$ and $E=\{{E}_{j}|j=1,2,...K\}$, respectively.
%Without loss of generality, 
We define the location of %the base station (BS),
the UAV relay, the eavesdropper, and the $i$th user in the 3D Cartesian coordinate system as %($x_{BS},y_{BS},z_{BS}$),
($x_u,y_u,z_u$) , ($x_j,y_j,z_j$), and ($x_i,y_i,z_i$), respectively.
%%The UAV flies at the % has a fixed 
%%altitude $z_u$, which may be chosen according to %considered as
%%the minimum height at which the UAV will have line of sight (LoS) links with the users in the cluster%, and where it would avoid buildings and obstacles from the air
%%~\cite{UAVLoS}. 
%%The UAV relay operation %along the cluster travel distance
%%will be executed over a finite duration $T$. 
For practicality, %we distribute the UAV relaying period over $N$ 
time slots are used to capture the different radio frames and statistical channel conditions, as well as the momentarily static position of the UAV, users, and eavesdroppers in a mobile setting. %, equally. 
Therefore, the UAV coordinates in time slot $n$ will be expressed as ($x_u[n], y_u[n], z_u[n]$).    

\vspace{-3mm}
\subsection{Communications Channel} The received signal at the UAV when ground user $V_i$ transmits signal $s_i$ with power $P_i$ is defined as
\vspace{-1 mm}
\begin{equation}
    {r _{u}} = \sqrt{P_i{g}_{iu}}s_i+n_i,
 %  \vspace{-2 mm}
\end{equation}
where ${g}_{iu}$ is the air-to-ground (A2G) channel power gain between user $V_i$ and the UAV, and $n_i$ denotes the %standard real 
additive white Gaussian noise (AWGN) of zero mean and $\sigma^2$ variance. % at the UAV. 
%As mentioned before, 
%A limited cluster size and flexible height of the UAV allows us to assume that the channel between ground users and UAV will be line of sight (LoS).
Measurements have shown that the LoS channel model is a good approximation in rural areas above 80 m~\cite{sky}. % links. 
The channel gain %${g}_{iu}[n]$ 
%in time slot $n$ 
has %is characterized by
a path loss exponent of two and can be written as
\vspace{-1 mm}
\begin{equation}
\small
\begin{aligned}
    {g}_{iu}[n] &=\beta_{0}d^{-2}_{iu}[n] \\
    &=\frac{\beta_0}{(x_i[n]-x_u[n])^2+(y_i[n]-y_u[n])^2+(z_i[n]-z_u[n])^2},
    \end{aligned}
\end{equation}
where $\beta_{0}$ is the channel gain at the reference point of $d_0$ = 1 m which is contingent on the transmitter and receiver antenna gains and the carrier frequency. Parameter $d_{iu}[n]$ represents the distance between the $i$th user and the UAV. % in the $n$th time slot. %, which is equal to $\sqrt{ (x_i[n]-x_u[n])^2+(y_i[n]-y_u[n])^2+(z_i[n]-H)^2}$. 
%The path loss exponent between ground users and the aerial relay is 2.

The capacity of the %legitimate 
channel between user $V_i$ and the UAV in time slot $n$ is given as 
\vspace{-1 mm}
\begin{equation}
%\small
%\begin{aligned}
    {C _{iu}[n]} = \log_2 \left( 1+\frac{{P_i}[n]{g}_{iu}[n]}{\sigma^2}\right) = \log_2 \left( 1+\frac{\lambda_0{P_i}[n]}{d^2_{iu}[n]}\right),
    %{(x_i[n]-x_u[n])^2+(y_i[n]-y_u[n])^2[n]+(z_i[n]-H)^2}
%\end{aligned}
\end{equation}
where $\lambda_0=\beta_0/\sigma^2$ is the signal-to-noise ratio (SNR) at the reference point. 

\subsection{Wiretap Channel} {Because the UAV is in the air and can often establish an LoS link with the BS, beamforming and PC can be effectively employed to focus the radio frequency (RF) transmission between the BS and UAV. %and reduce the radiated energy. 
On the other hand, a UE may have a single transmit antenna and, thus, its UL transmission will not be focused. Assuming ground eavesdroppers, here we therefore consider the user-UAV link being the dominant eavesdropping channel.}

The transmitted signal from user $V_i$ is overheard by eavesdropper ${E}_{j}$, which receives
\vspace{-2 mm}
\begin{equation}
    {r _{j}} = \sqrt{P_i{w}_{ij}}s_i+n_j,
%    \vspace{-2 mm}
\end{equation}
where $n_{j}$ denotes the %zero-mean 
AWGN %with a variance of $\sigma^2$ 
at the eavesdropper $E_j$. 
Because users and eavesdroppers are on the ground, the path loss exponent %between ground users and ground eavesdroppers 
is assumed to be four. The %${w}_{ij}$ is the 
ground-to-ground (G2G) channel gain between user $V_i$ and eavesdropper %node 
$E_j$ 
%the ${w}_{ij}$ at 
in time slot $n$ can then be modeled as
\vspace{-0.5 mm}
\begin{equation}
\small
\begin{aligned}
    %{w}_{ij}[n] &=\omega_{0}d^{-4}_{ij}[n]\\
    %=&\frac{\omega_0}{({(x_i[n]-x_j[n])^2+(y_i[n]-y_j[n])^2+(z_i[n]-z_j[n])^2)}^2},
    {w}_{ij}[n] &=\beta_{0}d^{-4}_{ij}[n]\\
    =&\frac{\beta_0}{({(x_i[n]-x_j[n])^2+(y_i[n]-y_j[n])^2+(z_i[n]-z_j[n])^2)}^2},
    \end{aligned}
\end{equation}
where %$\omega_0$ is the reference point's channel gain at $d_0$=1m,
$d_{ij}[n]$ is the distance between the $i$th user and the $j$th eavesdropper in the $n$th time slot. %, which is equal to $\sqrt{(x_i[n]-x_j[n])^2+(y_i[n]-y_j[n])^2+(z_i[n]-z_j[n])^2}$. 
The corresponding capacity is %of the wiretap channel between user $V_i$ and eavesdropper $E_j$ at time $n$ is then
\vspace{-2 mm}
\begin{equation}
%\small
%\begin{aligned}
    {C _{ij}[n]} = \log_2 \left( 1+\frac{{P_i}[n]{w}_{ij}[n]}{\sigma^2}\right) = \log_2 \left(
    1+\frac{\lambda_0{P_i}[n]}{d^{4}_{ij}[n]}\right).
    %1+\frac{\delta_0{P_i}[n]}{d^{4}_{ij}[n]})\right),
    %(x_i[n]-x_j[n])^2+(y_i[n]-y_j[n])^2[n]+(z_i[n]-z_j[n])^2
 %   \end{aligned}
\end{equation}
%where $\delta_0=\omega__0/\sigma^2$ is the signal-to-noise ratio(SNR) at the reference point. 

%This work considers the case of uncoordinated eavesdropping attacks. 
%between the available eavesdroppers. 
%%Therefore, if the transmission between the BS and $V_i$ has been interpreted successfully by any eavesdropper, the transmission will be exposed to other attackers. 
%As a result of that, 
%\textcolor{blue}{ We assume that the UE-AR links are subject to eavesdropping, whereas the BS-AR link can be more focused and is therefore less susceptible to eavesdropping.}

% As a result, the capacity of the eavesdropping channel is calculated as the maximum %achieved 
% capacity among the $K$ eavesdroppers:
% \vspace{-2.5 mm}
% \begin{equation}
%     {C _{ij}[n]} = \max\limits_{j \in E} C _{ij}[n] = \max\limits_{j \in E} \log_2
%     \left( 1+\frac{\lambda_0{P_i}[n]}{d^{4}_{ij}[n]}\right).
%     %\left( 1+\frac{\delta_0{P_i}[n]}{d^{4}_{ij}[n]}\right),
% \end{equation}
% \vspace{-10mm}
\subsection{Secrecy Capacity Metric} The common metric to evaluate the influence of eavesdropping and the performance of mitigation techniques is the secrecy capacity. The secrecy capacity is calculated as the difference between the legitimate and the wiretap channel capacities and corresponds to the rate at which no data will be decoded by the eavesdropper~\cite{YanWanGer2015}.  
The average secrecy capacity per cluster position over $N$ radio time slots with $M$ users and $E$ eavedroppers and one AR %of all users under attack \textcolor{blue}{per cluster center} 
is then obtained as %The C_sec^total will be computed for each  and user distribution  each movement.
\vspace{-2 mm}
\begin{equation}
    {C ^{Total}_{sec}} = \frac{1}{N}\sum\limits_{n=1}^N \Bigg[ 
    \sum\limits_{i=1}^M C _{iu}-\max\limits_{j \in E}\sum\limits_{i=1}^M C _{ij}[n] \Bigg]^+,
    \label{equ:totalsec}
\end{equation}
\vspace{4 mm}
where $[x]^+ \triangleq max(x,0)$%, \textcolor{blue}{and $N$ is the total equal length time slots of the flying period of the UAV per cluster center.} 
\section{Problem Formulation}
%\vspace{-3 mm}
The objective of this paper is maximizing the average secrecy rate $C ^{Total}_{sec}$ %~(\ref{equ:totalsec}) 
by optimizing the AR position and %user 
transmission power. % control. 

%The UAV position is a function of %subject to 
%the %maximum 
%radius of the mobile cluster $R$. 
In order to have a short distance to the affected users %difference 
with %and 
strong LoS connections, % with the effected UEs. %where the UAV position  
%the speed of the UAV where the maximum distance that the UAV can travel within each time slot is define as follow
%\vspace{-3 mm}
%\begin{equation}
%    {Q \limits_{max}[n]} = v\limits_{max}a_t,
%\end{equation}
%where $v_{max}$ is the maximum speed of the aerial relay and $a_t$ is the length of a time slot.
%Thus, 
the UAV position constraint is formulated as follows:
%\begin{equation}
%\begin{aligned}
%    (x_u[n+1]-&x_u[n])^2+(y_u[n+1]-y_u[n])^2 \leq Q^2,\\
%    &n=1,2,\cdots,N-1.
%    \end{aligned}
%\end{equation}
 \vspace{-1 mm}
\begin{equation}
    \sqrt{(x_u[n]-x_{center}[n])^2+(y_u[n]-y_{center}[n])^2} \leq R, \forall {n},
\end{equation}
%where $x_u[n+1]$ and $y_u[n+1]$ are the next time slot $x$th and $y$th UAV coordinates, respectively. While, $x_u[n]$ and $y_u[n]$ are the current time slot $x$th and $y$th UAV coordinates, respectively.
where $x_{center}[n]$ and $y_{center}[n]$ are the center coordinates of the mobile cluster at time slot $n$. 
We introduce %The user power is restricted by 
the average transmission %\hl{remove 'peak'}
power $\Bar{P}$ and the peak transmission %maximum 
power %\hl{Why? 'in each time slot'}
$ P_{max}$, where $ \Bar{P} \leq P_{max}$. The UL transmission power constraint for each user %in each time slot 
is %then 
 \vspace{-2 mm}
\begin{equation}
    \frac{1}{N}\sum\limits_{n=1}^N P[n] \leq \Bar{P},
    \label{P_avg_Max}
\end{equation}
\begin{equation}
   0 \leq P[n] \leq P_{max}, \forall {n}.
   \label{11}
\end{equation}

The optimization problem can now be written as
 \vspace{-2 mm}
\begin{equation}
\begin{aligned}
   (P1) : \max\limits_{x_u,y_u,P} \frac{1}{N}\sum\limits_{n=1}^N \Bigg[ %\frac{1}{M}
   \sum\limits_{i=1}^M  &\log_2  \left(1+\frac{\lambda_0{P_i}[n]}{d^2_{iu}[n]} \right)\\
   - \max\limits_{j \in E} \sum\limits_{i=1}^M&\log_2\
   \left(1+\frac{\lambda_0{P_i}[n]}{d^{4}_{ij}[n]}\right) \Bigg]^+ ,\\
   %\left(1+\frac{\delta_0{P_i}[n]}{d^{4}_{ij}[n]}\right) \Bigg]^+ ,\\
   s.t.\hspace{2 pt }(8),(9),(10).
   \end{aligned}
   \label{12}
\end{equation}
%\vspace{-4mm}
\section{%Optimization Problem 
Iterative Solution}
\label{Propo_Algo}

%This optimization problem will not have an optimum solution because 
The objective function is non-convex with respect the optimization parameters $x_u$, $y_u$, and $P$. In addition, the $[.]^+$ operator creates a non-smooth objective function. % at zero value. 
%As a result, we propose an iterative approach for solving it. 

$Lemma$ $1$: Problem $P1$ can be reformulated %altered % correspondingly 
as a new optimization problem $P2$ with the same optimal solution.
 \vspace{-2 mm}
\begin{equation}
\begin{aligned}
   (P2) : \max\limits_{x_u,y_u,P} \frac{1}{N}\sum\limits_{n=1}^N \Bigg[ %\frac{1}{M} 
   \sum\limits_{i=1}^M  &\log_2  \left(1+\frac{\lambda_0{P_i}[n]}{d^2_{iu}[n]} \right)\\
   -\max\limits_{j \in E} \sum\limits_{i=1}^M &\log_2
   \left(1+\frac{\lambda_0{P_i}[n]}{d^{4}_{ij}[n]}\right) \Bigg] ,\\
   %\left(1+\frac{\delta_0{P_i}[n]}{d^{4}_{ij}[n]}\right) \Bigg] ,\\
   s.t.\hspace{2 pt }(8),(9),(10).
   \end{aligned}
   \label{13}
\end{equation}
$Proof$ $of$ $Lemma$ $1$: %First, since that, the $[\upsilon]^+ \geq [\upsilon],  \forall{\upsilon}$ while we assume that 
Let $\xi_1^*$ and $\xi_2^*$ %are corresponding to 
be the optimal solutions for problems~(\ref{12}) and~(\ref{13}), respectively, and because $[\upsilon]^+ \geq [\upsilon]$, $\xi_1^* \geq \xi_2^*$. The optimal solution of~(\ref{12}) $\xi_1^*$ $=$ $(x^*_{u}, y^*_{u}, P^*)$, where $x^*_{u}=[x^*_{u}[1], x^*_{u}[2],...,x^*_{u}[N]]^\dagger, y^*_{u}=[y^*_{u}[1], y^*_{u}[2],...,y^*_{u}[N]]^\dagger$, and $ P^*=[P^*[1], P^*[2],...,P^*[N]]^\dagger$.
%Let $\tau(x_u[n],y_u[n],P[n])$ be defined as follow:
We define
%\begin{proof}
\begin{equation}
\begin{aligned}
\tau(&x_u[n],y_u[n],P[n])= \\
&\sum\limits_{i=1}^M  \log_2  \left(1+\frac{\lambda_0{P_i}[n]}{d^2_{iu}[n]} \right)
   -\max\limits_{j \in E} \sum\limits_{i=1}^M \log_2
   \left(1+\frac{\lambda_0{P_i}[n]}{d^{4}_{ij}[n]}\right).
   \end{aligned}
   \label{eq:xxx}
\end{equation}
%\end{proof}
Let $(\hat{x}_{u}, \hat{y}_{u}, \hat{P})$ represent the feasible values that are constructed while solving~(\ref{13}), where $\hat{x}_{u}=x^*_{u}, \hat{y}_{u}=y^*_{u}$ and $\hat{P[n]}$ can be computed as follows:
%\begin{proof}
\begin{equation}
\hat{P[n]}=
    \begin{cases}
       P^*[n] & \tau(x_u[n],y_u[n],P[n]) \geq 0,\\
      0 & \tau(x_u[n],y_u[n],P[n]) < 0,
    \end{cases}       
\end{equation}
%\end{proof}
%\textcolor{blue}{%Denote the constructed feasible solution of equation~(\ref{13}) defined as $\hat{\upsilon} $ at these equivalent points $(\hat{x}_{u}, \hat{y}_{u}, \hat{P})$. 
In other words, the UL transmission power is set to zero if (13) is negative. This will maximize (12), because the inner term of (12), i.e. (13), is 0 instead of being negative for a certain $n$, where an eavesdropper achieves a higher reception rate than the AR. This is equivalent to using the $[.]^+$ operator in (11). The constructed final solution $\hat{\upsilon} $ based on $(\hat{x}_{u}, \hat{y}_{u}, \hat{P})$ ensures that $\hat{\upsilon} = \xi_1^*$. Then, since $(\hat{x}_{u}, \hat{y}_{u}, \hat{P})$ represents a feasible solution for~(\ref{13})% over $n$
, consequently $\xi_2^* \geq \hat{\upsilon}$ and, therefore,  $\xi_2^* \geq \xi_1^*$. Since the premise at the beginning of the proof was that $\xi_1^* \geq \xi_2^*$, the equality holds, i.e. %As a result of that, we can conclude that the 
$\xi_1^* = \xi_2^*$, proving Lemma 1. %and the solution to (1.  
%}% and prove $Lemma$ $1$

%\textcolor{blue}{As proofed in $Lemma$ $1$, it is now the problem of finding the optimal solution of problem }$P2$ %will maintain
%\textcolor{blue}{that }still holds the restriction that the capacity of the legitimate user in each time slot must be higher than the capacity achieved at the eavesdropper similar to $P1$; otherwise, $P[n]$ is set to zero. %, thus, their optimal solution is the same. %Meanwhile, 
%still perplex getting the optimal solution for the $P2$ problem. Therefore,

The transformation from problem $P1$ to problem $P2$ eliminates the non-smoothness, but % enigma existed in the $P1$. 
$P2$ is still non-convex. We therefore %discuss using 
propose an iterative algorithm for solving $P2$. It is derived from the block coordinate descent strategy. In particular, we split the problem into two subproblems by separating the optimization variables in two blocks. The first subproblem optimizes the %users'
transmission powers with respect to a given UAV position. The second subproblem aims at finding the optimal UAV position for the given transmission powers. The solution of one subproblem is %used and 
applied to the other subproblem, %alternatively 
iteratively until convergence.

%\vspace{-20 mm}
\subsection{Subproblem P2.I: Transmission Power Optimization} %Under Provided Trajectory}

We apply $\log_2(z)= \ln (z)/\ln(2)$ to define the optimization subproblem $P2.I$ as 
%\vspace{-2 mm}
\begin{equation}
\begin{aligned}
   (P2.I) : \max\limits_{P} \frac{1}{N}\sum\limits_{n=1}^N \Bigg[ %\frac{1}{M} 
   \sum\limits_{i=1}^M &\ln \left(1+ \mu_n {P_i}[n]\right)\\
   - \max\limits_{j \in E} \sum\limits_{i=1}^M&\ln \left(1+\eta_n {P_i}[n]\right) \Bigg] ,\\
   %\vspace{-3 mm}
   s.t.\hspace{2 pt }(9),(10),
\label{P2.I}
%
%\frac{\delta_0}{d^{4}_{ij}[n]}
  \end{aligned}
\end{equation}
where 
\vspace{-4 mm}
\begin{equation}
    \mu_n = \frac{\lambda_0}{(x_i[n]-x_u[n])^2+(y_i[n]-y_u[n])^2+(z_i[n]-z_u[n])^2}
\end{equation}
%and 
\vspace{-4 mm}
\begin{equation}
    \eta_n =
    \frac{\lambda_0}{(x_i[n]-x_j[n])^2+(y_i[n]-y_j[n])^2+(z_i[n]-z_j[n])^2}.
    %\frac{\delta_0}{(x_i[n]-x_j[n])^2+(y_i[n]-y_j[n])^2+(z_i[n]-z_j[n])^2},
\end{equation}
%to obtain~(\ref{P2.I}).

Based on the findings of~\cite{P_optmalSolution}, the derivation of the optimal power allocation policy for subproblem~$P2.I$ can be easily determined for the case where $\mu_n \leq \eta_n$.  The objective function of~(\ref{P2.I}) with respect to ${P_i}[n]$ for this case is non-increasing and, therefore, the optimal transmission power is zero. For $\mu_n > \eta_n$, the objective function holds the non negative weighted sums(or integral) that preserves the concavity of the problem with respect to ${P_i}[n]$. Therefore, the optimal solution for it can be obtained by applying the Lagrangian maximization approach to get the solution of~(\ref{P2.I}), the following optimality condition has been found: 
\vspace{-1 mm}
%\begin{proof}
\begin{equation}
  \frac{\partial \tau(x_u[n],y_u[n],P[n])}{\partial {P}[n]} = \frac{\eta_n}{1+\eta_n{P}[n]}-\frac{\mu_n}{1+\mu_n{P}[n]}-\rho=0,
  \label{19}
  \vspace{-1.5 mm}
\end{equation}
%\end{proof}
where the power allocation policy at the transmitter $\Tilde{P}[n]$ that satisfies the optimality condition can be calculated as follows: 
%when $\mu_n > \eta_n$, while in case
\vspace{-2 mm}
\begin{equation}
  \Tilde{P}[n] =
    \begin{cases}
      min \left([{P^{\circ}}[n]]^+, P_{max}\right) & \mu_n \ge \eta_n,\\
      0 & \mu_n \leq \eta_n,
    \end{cases}       
\end{equation}
where 
\vspace{-2 mm}
\begin{equation}
   P^{\circ}[n] = \sqrt{\left(\frac{1}{2\eta_n}- \frac{1}{2\mu_n} \right)^2+ \frac{1}{\rho}\left( \frac{1}{\eta_n} - \frac{1}{\mu_n} \right)}-\frac{1}{2\eta_n}-\frac{1}{2\mu_n}.
\end{equation}
The constant $\rho \geq 0$ has been introduced to limit the %average
peak power according to %below the allowed maximum power in each time step that fulfill the constrain in
%\textcolor{blue}{the constrain in}
~(\ref{P_avg_Max}) and (\ref{11}), and it can be computed efficiently via a one-dimensional bisection search~\cite{rho}.
% \vspace{-1 mm}
 
\subsection{Subproblem P2.II: UAV Position Optimization}% Under Provided Transmission Power}
The second subproblem of problem $P2$ optimizes the UAV position. It can be formulated as %the optimization problem $P2$ will be transferred to this form, which will focus on , as follow
\vspace{-2 mm}
\begin{equation}
\begin{aligned}
   (P2.II) :  \max\limits_{x_u,y_u} \frac{1}{N}\sum\limits_{n=1}^N \Bigg[ %\frac{1}{M} 
   \sum\limits_{i=1}^M  &\log_2  \left(1+\frac{\grave{P_i}}{d^2_{iu}[n]} \right)\\
   -\max\limits_{j \in E}\sum\limits_{i=1}^M &\log_2
   \left(1+\frac{\grave{P_i}}{d^{4}_{ij}[n]}\right) \Bigg] ,\\
   %\left(1+\frac{\acute{P_i}}{d^{4}_{ij}[n]}\right) \Bigg] ,\\
  % \vspace{-5 mm}
   s.t.\hspace{2 pt }(8),
\label{P2.II}
%
%\frac{\delta_0}{d^{4}_{ij}[n]}
  \end{aligned}
\end{equation}
%\vspace{-1 mm}
where $\grave{P_i} = \lambda_0 {P_i}[n] $. %and $\acute{P_i} = \delta_0 {P_i}[n]$.

Here we %aim at 
optimize only the AR position. Hence, the focus is %on the $C_{iu}$ %first
%term of $P2.II$, %optimization problem regarding 
on maximizing the capacity between the ground users and the UAV. 
In order to solve this subproblem, which is a non-concave function with respect to $x_u$ and $y_u$, we add a slack variable $\psi=[\psi[1],\cdots,\psi[N]^\dagger$. The optimization problem can then be reformulated as 
%To solve the not concave optimization problem above with respect to $x_u, y_u$, we will add two slack variables $\tau=[\tau[1],\cdots,\tau[N]]^\dagger$ and $\psi=[\psi[1],\cdots,\psi[N]^\dagger$, respectivelly. Now, the new optimization problem can be shown as follow
% \vspace{-1 mm}
\begin{equation}
\begin{aligned}
   (P2.II) :  \max\limits_{x_u,y_u,\psi} \frac{1}{N}\sum\limits_{n=1}^N \Bigg[ %\frac{1}{M} 
   \sum\limits_{i=1}^M &\log_2  \left(1+\frac{\grave{P_i}}{{\psi}[n]} \right) \\
   -\max\limits_{j \in E}\sum\limits_{i=1}^M &\log_2
   \left(1+\frac{\grave{P_i}}{d^{4}_{ij}[n]}\right) \Bigg] ,\\
  \end{aligned}
     \tag{22.a}
\label{P2.II.a}
\end{equation}
%\begin{equation}
%\begin{aligned}
%   (P2.II) :  \max\limits_{x_u,y_u,\tau,\psi} \frac{1}{N}\sum\limits_{n=1}^N \Bigg[ \frac{1}{M} \sum\limits_{i=1}^M  &\ln  \left(1+\frac{\grave{P_i}}{{\psi}[n]} \right) \\
%   -\max\limits_{j \in E} &\ln \left(1+\frac{\acute{P_i}}{\tau[n]}\right) \Bigg] ,\\
%  \end{aligned}
%     \tag{21.a}
%\label{P2.II.a}
%\end{equation}
\vspace{-0.5 mm}
   s.t.\hspace{2 pt }
%   \begin{equation}
%   \begin{aligned}
%    \tau[n]-\bigg((x_i[n]-x_j[n])^2+&(y_i[n]-y_j[n])^2\\
%    +&(z_i[n]-z_j[n])^2\bigg)^2 \leq 0, \forall {n}.
%  \end{aligned}
%  \tag{21.b}
%\end{equation}
%\vspace{-1 mm}
\begin{equation}
   \begin{aligned}
    \psi[n]-x^2_i[n]&+2x_i[n]x_u[n]-x^2_u[n]-y^2_i[n]+2y_i[n]y_u[n]\\&-y^2_u[n]
    -z^2_i[n]+2z_i[n]H-H^2 \leq 0, \forall {n},
  \end{aligned}
\tag{22.b}
\label{P2.II.a_st1}
\end{equation}
%\begin{equation}
%   \begin{aligned}
%    \psi[n]-\bigg((x_i[n]-x_u[n])^2+&(y_i[n]-y_u[n])^2\\
%    +&(z_i[n]-H)^2\bigg) \leq 0, \forall {n}.
%  \end{aligned}
%\tag{21.c}
%\end{equation}
% \vspace{-5 mm}
\begin{equation}
    \psi[n]\geq 0, \forall {n}. \tag{22.c}
%    \vspace{-1 mm}
\end{equation}
 %\hspace{35 mm}(9).
 %\vspace{-1 mm}

The equality %equations 
in (22.b) and (22.c) shall be satisfied for the optimal solution. % along with its constraint~(\ref{P2.II.a_st1}). 
If this is not the case, then the slack variable $\psi[n]$ can be increased until an improvement in the objective function is achieved. 
%On account of that, 
As a result, we can consider that the ideal result of (22) will be the same for~(\ref{P2.II}).  
  \begin{figure*}[ht]
     \centering
     \vspace{-3 mm}
     \includegraphics[width=1\textwidth]{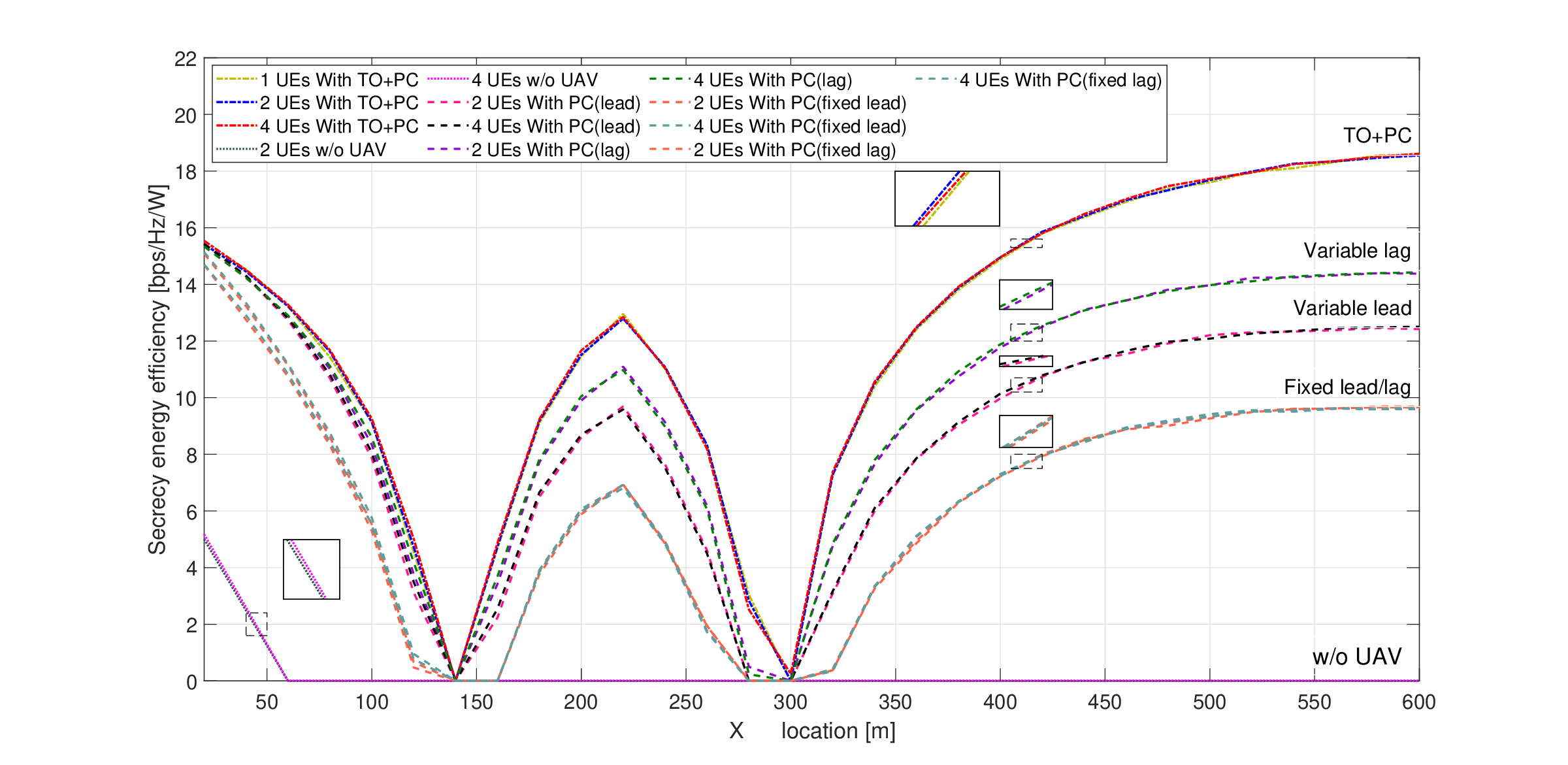}
     \vspace{-15 pt}
     \caption{SEE %Secrecy energy efficiency 
     for multiple mobile UEs under attack by multiple eavesdroppers and different mitigation strategies.%proposed algorithm against leading or lagging UAV trajectories.
     }
     \vspace{-1 mm}
     \label{fig:sec_result2}
 \end{figure*}
 %\vspace{-4 mm}

In the process of finding the optimal solution for~(\ref{P2.II.a}), we must consider the terms $-x^2_u[n]$ and $-y^2_u[n]$ that conceive the non-convexity. %and harden finding an optimal solution for problem~(\ref{P2.II.a}).
Therefore, we employ an iterative algorithm that uses the successive convex optimization method for developing an approximate solution for~(\ref{P2.II.a}) by maximizing the concave lower bound of the objective function within the convex feasibility region. %\textcolor{blue}{The approximate solution %to the problem 
%can be obtained by reaching the maximum concave lower bound of the objective function within the convex feasibility region.} % limit.}

First, we introduce $x_{u(fea)} \triangleq  [x_{u(fea)}[1],\cdots,x_{u(fea)}[N]]^\dagger $, $y_{u(fea)} \triangleq  [y_{u(fea)}[1],\cdots,y_{u(fea)}[N]]^\dagger $, and $\psi_{(fea)}$ $\triangleq$  $[\psi_{(fea)}[1],\cdots,\psi_{(fea)}[N]]^\dagger$, where this initial point $(x_{u(fea)}, y_{u(fea)}, \psi_{(fea)})$ is feasible to~(22). %, which is feasible for equations~(\ref{P2.II.a_st1}) and~(\ref{P2.II.a}). 
By taking into consideration that $-x^2_u[n]$ and $-y^2_u[n]$ are concave functions, their first-order Taylor expansions at $x_{u(fea)}$, $y_{u(fea)}$, and $\psi_{(fea)}$ are global over-estimators. Therefore, the next step will be calculating the first-order Taylor expansions at %these points
$x_{u(fea)}$, $y_{u(fea)}$, and $\psi_{(fea)}$ for the following terms $\sum\limits_{i=1}^M \log_2   \left(1+\frac{\grave{P_i}}{{\psi}[n]} \right)$, $-x^2_u[n]$, and $-y^2_u[n]$, respectively: 
%\vspace{-1 mm}
%\begin{proof}
\begin{equation}
\begin{aligned}
    \sum\limits_{i=1}^M \log_2  \left(1+\frac{\grave{P_i}}{{\psi}[n]} \right) &\geq \sum\limits_{i=1}^M \log_2  \left(1+\frac{\grave{P_i}}{{\psi_{(fea)}}[n]} \right)\\
    &-\frac{\grave{P_i}({\psi}[n]-{\psi_{(fea)}})}{ln2({\psi}^2_{(fea)}[n]+\grave{P_i}{\psi_{(fea)}})},
    \end{aligned}
        \tag{23}
\end{equation}
%\end{proof}
 \vspace{-3 mm}
\begin{equation}
    -x^2_u[n] \leq x^2_{u(fea)}[n]-2x_{u(fea)}[n]x_u[n],
    \tag{24}
\end{equation}
 \vspace{-5 mm}
\begin{equation}
    -y^2_u[n] \leq y^2_{u(fea)}[n]-2y_{u(fea)}[n]y_u[n].
    \tag{25}
\end{equation}
%\vspace{-2.5 mm}
As a result, (\ref{P2.II.a}) and its constraint~(\ref{P2.II.a_st1}) can be reformulated as %follows 
\vspace{-3 mm}
\begin{equation}
\begin{aligned}
   (P2.II) :  \max\limits_{x_u,y_u,\psi} \frac{1}{N}\sum\limits_{n=1}^N \Bigg[ %\frac{1}{M}
   \sum\limits_{i=1}^M &-\frac{\grave{P_i}({\psi}[n])}{ln2({\psi}^2_{(fea)}[n]+\grave{P_i}{\psi_{(fea)}})} \\
   -\max\limits_{j \in E}\sum\limits_{i=1}^M &\log_2
   \left(1+\frac{\grave{P_i}}{d^{4}_{ij}[n]}\right) \Bigg] ,
  \end{aligned}
     \tag{26.a}
\label{P2.II.solution}
\end{equation}
 \vspace{-2 mm}
   s.t.\hspace{2 pt }
   \begin{equation}
   \begin{aligned}
    \psi[n]&-x^2_i[n]+2x_i[n]x_u[n]+x^2_{u(fea)}[n]-2x_{u(fea)}[n]x_u[n]\\&-y^2_i[n]+2y_i[n]y_u[n]+y^2_{u(fea)}[n]-2y_{u(fea)}[n]y_u[n]
    \\&-z^2_i[n]+2z_i[n]H-H^2 \leq 0, \forall {n},
  \end{aligned}
\tag{26.b}
\label{P2.II.final}
  %  \vspace{-2 mm}
\end{equation}
 %\hspace{35 mm}(9).
 \vspace{-4 mm}
\begin{equation}
    \psi[n]\geq 0, \forall {n}. \tag{26.c}
    \vspace{-1 mm}
\end{equation}
This final version of the UAV position optimization problem %is~(23) 
is concave %optimization problem 
with a convex feasibility region. 
Therefore, the solution for it can be obtained by using the interior point method~\cite{optimization_book}.  

Since the first constraint of problem (26) covers that of problem (22), the feasible solution to %problem 
(22) can be found by solving (26). 
In addition, it is guaranteed that applying the solution of (26) to the objective function of (22) will not result in a lower value when using $(x_{u(fea)}, y_{u(fea)}, \psi_{(fea)})$. %This is so because ($x_{u(fea)}, y_{u(fea)}$) are the only points at which the lower bound and objective function of problem (20) can be equal \hl{to what?}. However, the solution to problem (23) aims at maximizing the lower bound of the objective function of problem (20).    
%%This is so because at the above point %($x_{u(fea)}, y_{u(fea)}$) 
%%the lower bound and the objective function of (22) are equal.  %provides the equality of  %solution 
%for %the objective function of problem 
%(22). }
%However, by finding a solution to %problem 
%(27), the lower bound of %the objective function of problem 
%(23) will be maximized.  
%\vspace{-4 mm}

%\vspace{-3 mm}
\begin{figure}[!t]
 \removelatexerror
  \begin{minipage}{0.95\linewidth}
%  \begin{adjustwidth}{1em}{}
\begin{algorithm}[H]
\footnotesize
\SetAlgoLined
%\begin{enumerate}[ itemsep=-10pt, topsep=2pt]
    %\item 

    Initialize $P^{(0)}$, $x_u^{(0)}$, $y_u^{(0)}$, and the initial slack variable $\psi^{(0)}$ to feasible values and set $j=0$;
    %\item 
    
    Calculate $C^{(0)}=f_{P2}(P^{(0)}, x_u^{(0)}, y_u^{(0)})$ according to~(\ref{13});
    
    \While{ $\frac{C^{(j)}-C^{(j-1)}}{C^{(j)}} \leq \chi$}
  {
        With given $P^{(j-1)}$, set the new feasible points $x^{(j)}_{u(fea)}=x^{(j-1)}_u$, $y^{(j)}_{u(fea)}=y^{(j-1)}_u$, and $\psi^{(j)}_{(fea)}=\psi^{(j-1)}$;
        
        Calculate the new trajectory update $(x^{(j)}_u,y^{(j)}_u)$ and the slack variable $\psi^{(j)}$  according to~(26);
        
        Solve equation (19) with the obtained trajectory $(x^{(j)}_u,y^{(j)}_u)$ to calculate the updated transmission powers $P^{(j)}$;
        
        Calculate the new $C^{(j)}=f_{P2}(P^{(j)}, x_u^{(j)}, y_u^{(j)})$ according to~(\ref{13});
   		
  		$j \Leftarrow j+1 $
   		
  }
  \KwResult{Optimal $P$, $x_u$ and $y_u$ }
    %\textbf{until} $\Vert \nabla_{R} \xi(\vartheta) \Vert_2 \leq \chi$\\
 \caption{\small Proposed algorithm for optimizing the transmission power and UAV position.}
 \label{Algorithm1}
\end{algorithm}
% \end{adjustwidth}
  \end{minipage}
    \end{figure}

  \subsection{Overall Algorithm}
Algorithm~1 presents the details of the proposed optimization method, where the block coordinate descent method is deployed for finding a suboptimal solution to Problem~$P1$ in an iterative manner. Parameter $\chi$ is the threshold indicator of the convergence accuracy.%, and $J$ is the %total number of 
%iteration variable of the iterative algorithm.} 

The complexity of Algorithm~1 can be obtained from the pseudocode as $\mathcal{O}(J((2M+E)N)^{3.5})$, where $J$, $M$, $E$, and $N$ are the number of iterations, users, eavesdroppers, and radio time slots, respectively. The inner term $(2M+E)N$ come from (7), the $((2M+E)N)^{3.5}$ is the algorithm complexity per iteration for obtaining the new location and PC parameters, and $J((2M+E)N)^{3.5}$ is the total complexity of the iterative algorithm. 
This complexity is polynomial and feasible for practical use. % with a reasonable time consumption compared to any exponential complexity. 

%\textcolor{blue}{
The convergence of Algorithm 1 is proven as follows. First we note that in iteration $j \geq 1$,  %after the executions steps 4 to 7 that 
the objective function~(\ref{13}) is non-decreasing: 
 %  \begin{equation}
% \vspace{-1 mm}
 %\begin{proof}
 \begin{minipage}{1\linewidth}
 \vspace{-2 mm}
   \begin{align}
   \Omega (&x^{(j-1)}_u,y^{(j-1)}_u,P^{(j-1)})\tag*{} \\
   &=\Upsilon(x^{(j-1)}_u,y^{(j-1)}_u,\psi^{(j-1)},P^{(j-1)})\tag{27.a}\\
   &=\Gamma(x^{(j-1)}_u,y^{(j-1)}_u,\psi^{(j-1)},P^{(j-1)})\tag{27.b}\\
   &\leq \Gamma(x^{(j)}_u,y^{(j)}_u,\psi^{(j)},P^{(j-1)})\tag{27.c}\\
   &\leq \Upsilon(x^{(j)}_u,y^{(j)}_u,\psi^{(j)},P^{(j-1)})\tag{27.d}\\
   &=\Omega(x^{(j)}_u,y^{(j)}_u,P^{(j-1)}),\tag{27.e}
  \end{align}
   \end{minipage}
   %\vspace{1 mm}
 %  \end{proof}
%\label{P2.II.final}
%\end{equation}
where $\Omega,\Upsilon$, and $\Gamma$ correspond to the the objective functions (12), (22), and (26), respectively. \textbf{In steps (4-5)}, (21) and (22) share the exact optimal value and solution of $(x_u,y_u)$; hence, (27.a) and (27.e) hold. The feasible values $(x_{u(fea)}$, $y_{u(fea)}$, $\psi_{(fea)})=(x^{(j-1)}_u,y^{(j-1)}_u,\psi^{(j-1)})$  make the first-order Taylor expansions in {(23), (24), and (25)} tight; therefore, (27.b) holds. The optimal solution to (26) is $(x^{(j)}_u,y^{(j)}_u,\psi^{(j)})$ and enforces %the hold on 
(27.c). Since $\Gamma$ is a lower bound of (22), (27.d) holds. \textbf{Steps (6-7)} calculate the new transmission powers and update the solution: 
%\begin{proof}
\begin{equation}
    \Omega (x^{(j)}_u,y^{(j)}_u,P^{(j-1)}) \leq \Omega (x^{(j)}_u,y^{(j)}_u,P^{(j)}).
    \vspace{-2 mm}
    \tag{28}
\end{equation}
%\end{proof}
We combine (28) and (27.e) and obtain
%\begin{proof}
\begin{equation}
    \Omega (x^{(j-1)}_u,y^{(j-1)}_u,P^{(j-1)}) \leq \Omega (x^{(j)}_u,y^{(j)}_u,P^{(j)}).
    \vspace{-2 mm}
    \tag{29}
\end{equation}
%\end{proof}
From the above, we have proven that $\Omega$ is non-decreasing and upper-bounded by a finite value; therefore, Algorithm 1 is guaranteed to converge.
%\newline

%     \begin{minipage}{\linewidth}
%   \begin{algorithm}[H]
  
%     \caption{Proposed algorithm for optimizing the transmission power and UAV Position.}\label{euclid}
%     \begin{algorithmic}[1]
%      % \Procedure{Euclid}{$a,b$}\Comment{The g.c.d. of a and b}
%         \State  Initialize $P^{(0)}$, $x_u^{(0)}$, $y_u^{(0)}$, and the initial slack variable $\psi^{(0)}$ to feasible values and set $j=0$;
        
%         \State Calculate the $C^{(0)}=f_{P2}(P^{(0)}, x_u^{(0)}, y_u^{(0)})$ according to~(\ref{13});
        
%         \While{ $\frac{C^{(j)}-C^{(j-1)}}{C^{(j)}} \leq \chi$}
        
%           \State With given $P^{(j-1)}$, set the new feasible points $x^{(j)}_{u(fea)}=x^{(j-1)}_u$, $y^{(j)}_{u(fea)}=y^{(j-1)}_u$, and $\psi^{(j)}_{(fea)}=\psi^{(j-1)}$;
          
%           \State Calculate the new trajectory update $(x^{(j)}_u,y^{(j)}_u)$ and the slack variable $\psi^{(j)}$  according to~(27);
          
%           \State Solve equation (20) with the obtained trajectory points $(x^{(j)}_u,y^{(j)}_u)$ to calculate the updated transmission power $P^{(j)}$;
          
%         \EndWhile\label{euclidendwhile}
        
%         \State $j \Leftarrow j+1 $
%       %\EndProcedure
%     \end{algorithmic}
%   \end{algorithm}
%   \end{minipage}
\vspace{-2 mm}

\section{Numerical Analysis and Discussion}
\vspace{-1 mm}
\label{sec:results}

Numerical results are provided to evaluate the performance of the proposed technique. %to enhance the secrecy of multiple UEs under smart eavesdropping attack.
%%Considering a cluster of ground users that moves along the x axis. 
%with a fixed step size $dx$ where $UE_{X_{t+1}} = UE_{X_{t}} + dx$. 
Without loss of generality, the user cluster contains one, two or %up to four
four single-antenna UEs that are randomly distributed in an area with a 30 m radius. The simulation scenario contains two fixed eavesdroppers that are both able to wiretap the multiuser UL channel. %% choose %apply a smart strategy based on defining 
%%the closet UEs within the cluster to wiretap their UL transmissions. 
%%The %applied %smart 
%%eavesdroppers thus attack the most vulnerable users. 
%, those that are close or have a good channel links, by %strategy 
%leveraging %rises the probability of having 
%LoS links between each eavesdropper and UE under attack. 
%%This maximizes the wiretapping rates at the eavesdroppers.  %\textcolor{red}{and is the worst case for the users???} \textcolor{blue}{This is a smaller distance to eavesdropper means its ability to interpret the data increases as it has a higher SNR}. 

A UAV relay flies at a constant height of 80 m and is equipped with an omidirectional antenna to interface with the ground UEs. Its location is optimized according to Algorithm~1, starting at the cluster center. More precisely, the UAV follows the user cluster and %for a given location of the user cluster, %starts its operation from its initial point and travels closer to the ground user cluster. At each step, the UAV finds 
%the UAV 
recalculates its position as a function of the determined UE transmission powers in an iterative optimization process that optimizes power and position. %, applying Algorithm~1.% in Section~\ref{Propo_Algo}. 
%The AR is at a fixed height of 80 m. % and follows the user. 

%\textcolor{red}{\hl{10 m radius seems small. What is this choice based on?}} \textcolor{blue}{I have tried different sizes of cluster and I think the same behaviour apply for most of the cases from 10 to 80 m, So I think this was the last thing we can change it with confident to higher value } %, unless otherwise is specified. 

In order to show the effectiveness of the proposed scheme, we evaluate other UAV positions. %ing mechanisms for relaying the UL data. 
%%\textcolor{blue}{One option is the \textit{leading} AR where the UAV} %first approach is denoted as  
%%moves at a higher velocity than the ground user cluster. 
%\textcolor{red}{Are the velocities of the UE cluster and UAVs constant? Are the UEs redistributed within the cluster at each position or remain in their relative positions when moving forward? What are the channel models and do you consider Doppler due to mobility? Can we include the channel model in the Table.} \textcolor{blue}{In the proposed algorithm, the UE and UAV moves with the same velocity while in leading higher velocity and lagging lower. No, the UE redistributed once at beginning of the trip and then moves with constant speed.}% and reaches the next position before the UEs. 
%%\textcolor{blue}{The other is the \textit{lagging} AR where the UAV} travels at a lower velocity than the ground users. 
Specifically, we consider %For a one dimensional mo model, the %In the simulation scenarios, we apply two mechanisms for the
leading and lagging UAV positions: Variable leading and lagging here refers to the UAV flying at 2x and 0.5x the speed of the cluster movement, respectively, %have been selected as examples of the leading and lagging AR scenarios, 
whereas fixed leading/ lagging refers to the UAV being off by +/-50 m with respect to the center of the cluster. %that makes the UAV has lagging position than UEs. 
These baselines %techniques hence do not 
implement the proposed PC mechanism without the trajectory optimization (TO). %In the case of not applying the PC strategy,} %without power control (PC), and 
%the transmission powers \textcolor{blue}{will be} equally distributed in each time slot as $P[n] = \bar P, \forall{n}$. 
Table~\ref{tab:parameters} provides the simulation parameters. 
\begin{table}[]
%\fontsize{12}{10}\selectfont
\fontsize{4}{8}\selectfont
\centering
\caption{Simulation parameters.}
\label{tab:parameters}
\resizebox{0.35\textwidth}{!}
{%
\begin{tabular}{|c|c|}
\hline
\begin{tabular}[c]{@{}c@{}}\textbf{Simulation  parameter}\end{tabular}     & \textbf{Value}                              \\ \hline
\begin{tabular}[c]{@{}c@{}}Cluster initial center position\end{tabular} & (10, 0, 1.5) m                          \\ \hline
\begin{tabular}[c]{@{}c@{}}Cluster radius\end{tabular} & 30 m                            \\ \hline
\begin{tabular}[c]{@{}c@{}} 1st eavesdropper position\end{tabular}    & (140, 0, 1.5) m                          \\ \hline
\begin{tabular}[c]{@{}c@{}} 2nd eavesdropper position\end{tabular}    & (300, 0, 1.5) m                          \\ \hline
\begin{tabular}[c]{@{}c@{}}UAV relay initial position\end{tabular}       & (10, 0, 80) m \\ \hline

%Noise variance                                                     & 10^{\text{-12} }       \\ \hline
%\begin{tabular}[c]{@{}c@{}}
%SNR at the reference point\end{tabular}                                                      & 90 dB       \\ \hline
%\begin{tabular}[c]{@{}c@{}}dx    \end{tabular}        & 20 m                               \\ \hline
%M                                                          & 5 users                              \\ \hline
\begin{tabular}[c]{@{}c@{}}Leading \& lagging transmit powers\end{tabular}                                                          & 1 W                              \\ \hline
Variable leading \& lagging factor          
  & 2x \& 0.5x                \\ \hline
Fixed leading \& lagging distance          
  & +/-50 m                 \\ \hline
  
% Bandwidth                                                          & 10 MHz                              \\ \hline
\end{tabular}%
}
\vspace{-5mm}
\end{table}

Fig.~\ref{fig:sec_result2} illustrates the SEE results at the different locations along the traveled path of the user cluster. % for different schemes with different number of served users. %\textcolor{blue}{The simulation scenarios embrace a leading and lagging cases for the flying speed of the UAV with respect to the speed of mobile user. A leading and }
The results show the cases where the eavesdroppers attack %selects the number of victims as
a cluster of one, two, or four mobile UEs. 

The curves show that the SEE performances when applying the proposed joint %position
optimization method %and power control technique 
are nearly identical for a single, two, or four UEs that are served by the AR. 
Overall, the proposed technique shows outstanding performance in terms of SEE when compared to no relaying and superior performance when compared to other UAV positions. 
Additional simulations with more UEs in the cluster show similar results. 
This verifies the scalability of the AR assisted technique as an effective mitigation strategy against eavesdropping. 

%For comparison, we study three benchmark system designs:
% \begin{figure}[ht]
% \label{fig:results3}
% %  \subfloat[]{
% %	\begin{minipage}[c][0.55\width]{
% %	   0.55\textwidth}
% %	   \centering
% %	   \includegraphics[width=0.7\textwidth]{Figures/Secrecy_2UEs_UAV_TO_Iterations_TextUpdate.eps}
% %	    \label{fig:results_ite2}
% %	\end{minipage}}
% % \hfill 	
% %  \subfloat[]
% %  {
% %	\begin{minipage}[c][0.55\width]{
% %	   0.4\textwidth}
% 	   \centering
% 	   \includegraphics[width=0.32\textwidth]{Figures/Secrecy_4UEs_UAV_TO_Iterations_TextUpdate.eps}
% 	    %\label{fig:results_ite4}
% %	\end{minipage}}
% 	\vspace{-2 pt}
% \caption{Trajectory optimization convergence for %(a) two UEs, and (b) 
% four UEs and two UAV heights.
% }
% \vspace{-3mm}
% \end{figure}

%\begin{figure*}[ht]
\begin{figure}[ht]
 % \subfloat[]{
%	\begin{minipage}[c][0.65\width]{
%	   0.34\textwidth}
	   \centering
	   \includegraphics[width=0.40\textwidth]{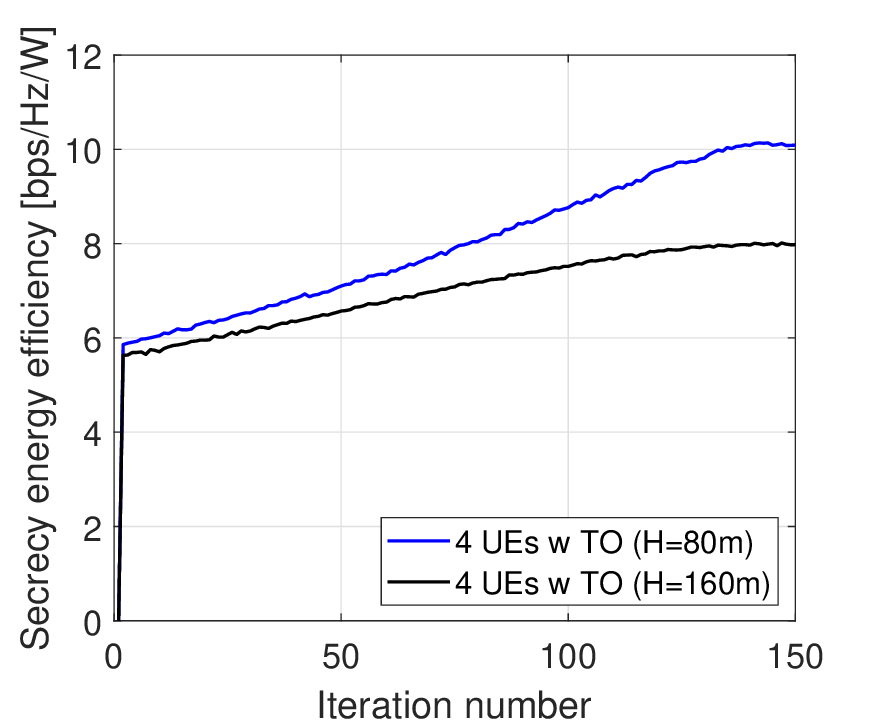}

	    \caption{Trajectory optimization convergence for %(a) two UEs, and (b) 
	    four UEs and two UAV heights.}
	    	    \label{fig:results3}
\end{figure}
%	\end{minipage}}
 %\hfill 
%  \subfloat[]{
%	\begin{minipage}[c][0.65\width]{
%	   0.34\textwidth}
\begin{figure}[ht]
	   \centering
	   \includegraphics[width=0.40\textwidth]{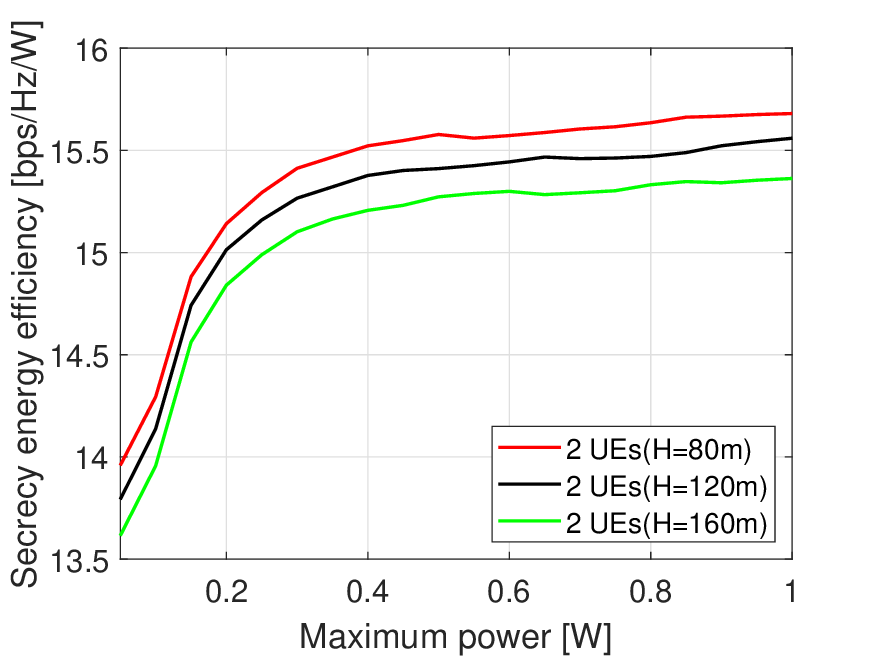}

	    \caption{SSE for two UEs as a function of the maximum transmission power and UAV height.}
	    	    \label{fig:results_Pow2}
	    \end{figure}
%	\end{minipage}}
 %\hfill 	
%  \subfloat[]
%  {
%	\begin{minipage}[c][0.65\width]{
%	   0.34\textwidth}
\begin{figure}[ht]
	   \centering
	   \includegraphics[width=0.40\textwidth]{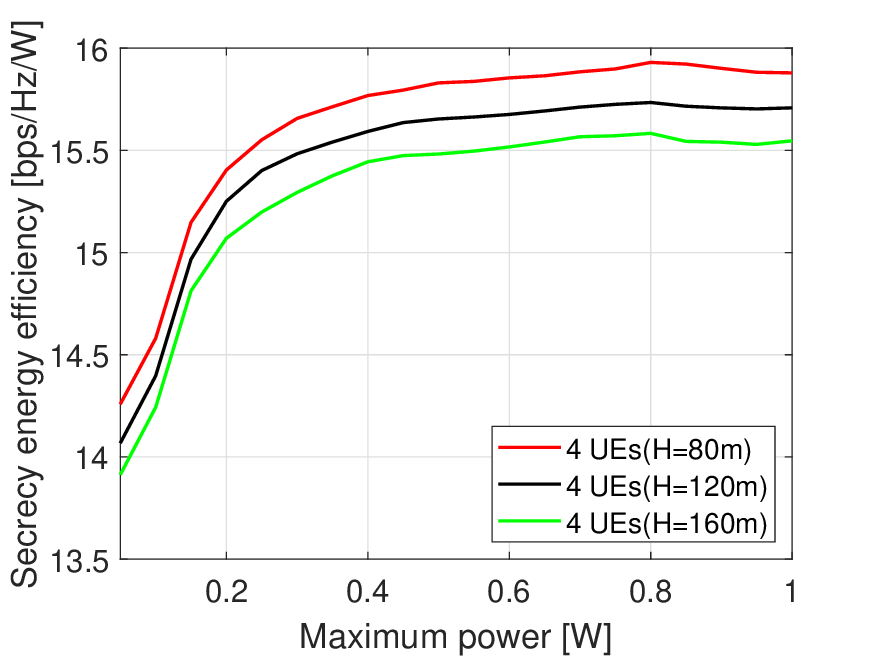}

%	\end{minipage}}
	\vspace{-2 pt}
\caption{SSE for four UEs as a function of the maximum transmission power and UAV height.
}
	    \label{fig:results_Pow4}
\end{figure}
 %\textcolor{red}{\hl{How is the eavesdropping rate calculated/combined among the two eavesdroppers?}}\textcolor{blue}{finding the max between both eaves for each time slot} 
%%In addition to that, the figure contains the multiple users cases (two and four) SEE while using the leading and lagging techniques for the purpose of comparison. 
 %the other benchmark schemes regarding the higher SEE achieved.

Fig.~\ref{fig:results3} shows %compares 
the convergence behavior of the %used
iterative algorithm for %two and 
four UEs under attack and two UAV heights. 
The corresponding curves for one and two UEs in the cluster are nearly identical and are therefore omitted. %Fig.~\ref{fig:results_ite2} shows improvement in the SEE for two UEs case under different two heights of the UAV relay. For both cases, 
The initial SEE is nearly 6~bps/Hz/W and after 130 iterations, the algorithm %comparatively 
converges at 10 and 8~bps/Hz/W for  %for the two UAV heights of 
UAV heights of 80 and 160~m, 
respectively. %The algorithm showed almost the same behaviour while four UEs under eavesdropping attack as shown in Fig.~\ref{fig:results_ite4}.

Figs.~\ref{fig:results_Pow2} and~\ref{fig:results_Pow4} present the %variation of the 
SEE performance as a function of the maximum UL transmission power $P_{max}$ %of ground users 
for three UAV heights. 
{We chose a range that represents %the maximum transmission power from the 
low-power IoT devices %at 0.05 W 
to high power public safety UEs.} %devices' level around 1 W.} 
Both results--- Fig.~\ref{fig:results_Pow2} and Fig.~\ref{fig:results_Pow4}---show that the SEE rapidly increases and reaches a steady state at about 400~mW. % UE transmission power. 
Note that the SEE performances for 200~mW, which legacy %today's 
4G phones use, %are capable of,
and 400~mW differ by only 0.25-0.5~bps/Hz/W.  %the beginning then the acceleration of the SEE becomes smaller until eventually reaches a steady value. 
The reason behind the monotonic behavior is that the SEE is a non-decreasing function of the maximum transmission power. 
%%From Fig.~\ref{fig:results_Pow2} the SEE of two users when the UAV flies at 80, 120, and 160 m becomes nearly constant when the transmit power exceed 0.4 W. 
%%The SEE behaviour of four users in the same conditions have been simulated and the results, depicted in Fig.~\ref{fig:results_Pow4}, show equivalent behaviour. % that prove %. The obtained results proved 
%%That is, the SEE does not necessarily improve by increasing the transmit power. % and anm optimal point can be found that balances energy efficiency and secrecy. 
Therefore, carefully regulating the transmission power can achieve energy efficient communications and attack mitigation. 

The obtained results %shown in Figs.~2 and 3 
confirm that combining the optimization of the transmission powers of ground users and the AR trajectory can significantly improve %(1) 
the UE power efficiency and %(2) 
the SEE by almost %17\% 
50\% in the tight attack zone between the two eavesdroppers. % and by over 30\% when the cluster travels away from the attack zone.

%Note that the leading and lagging UAVs transmit at 1~W so that the SEE differences in Fig.~2 are observed due to the UAV position. % technique.     
%An exciting result reveals that  EE gain is increased significantly by employing the proposed algorithm as compared to the benchmarks.

%  \begin{figure}[H]
%      \centering
%      \includegraphics[width=0.5\textwidth]{Figures/Secrecy2UEsUAVTOIterations.eps}
%      \caption{Trajectory optimization conversion behavior of 2 UEs at different UAV's height.}
%      \label{fig:results_ite2}
%  \end{figure}
%   \begin{figure}[H]
%      \centering
%      \includegraphics[width=0.5\textwidth]{Figures/Secrecy_4UEs_UAV_TO_Iterations.eps}
%      \caption{Trajectory optimization conversion behavior of 4 UEs at different UAV's height.}
%      \label{fig:results_ite4}
%  \end{figure}
%   \begin{figure}[H]
%      \centering
%      \includegraphics[width=0.5\textwidth]{Figures/Secrecy_2UEs_UAV_Power.eps}
%      \caption{Secrecy efficiency rate of 2 UEs versus the maximum transmission power.}
%      \label{fig:results_Pow2}
%  \end{figure}
%   \begin{figure}[H]
%      \centering
%      \includegraphics[width=0.5\textwidth]{Figures/Secrecy_4UEs_UAV_Power.eps}
%      \caption{Secrecy efficiency rate of 4 UEs versus the maximum transmission power.}
%      \label{fig:results_Pow4}
%  \end{figure}

\section{Conclusions}
\vspace{-1 mm}
\label{sec:conclusions}
UAV-based ARs have been proposed as a mitigation technique against eavesdropping because of their high mobility and maneuverability in the 3D space, LoS links with ground nodes, and their low cost. 
%A lot of current 
Research has considered %its usage in this 
ARs for this purpose, but has not analyzed the scalability of a UAV for supporting multiple mobile users. %its performance while serving multiple nodes 
Because of its importance for practical deployments, %was an important and growing research question.
%Therefore, in 
this paper has %therefore %we have 
studied this aspect and formulated %e scalability performance metric of the UAV relay solution enhanced
and solved a %n optimization
problem that optimizes UE transmission powers and UAV positioning. %the aerial relay position and ground nodes transmission power. 
The obtained results and %extracted
key findings show that the use of a single UAV as an AR is an efficient and scalable mitigation scheme for multiple mobile users against distributed eavesdroppers. %whose location is unknown.  %attacks. 
%For future work, we are planing to investigate the performance of the UAV relay assisted by Reconfigurable intelligent surface (RIS) communication to boost the spectral efficiency of the system.
Future work will assess the secrecy of the BS-UAV link in different settings as well as analyze the relation of the cluster radius, number of UEs, and adaptive UAV altitudes with their corresponding channel models
%and LoS implication 
on the optimal AR deployment. We will also validate the proposed techniques on suitable research platforms, such as the Aerial Experimentation and Research Platform for Advanced Wireless (AERPAW)~\cite{marojevic2020advanced}.

%for assessing

\balance

\bibliographystyle{IEEEtran}
\bibliography{REFs}

\end{document}